\documentclass[runningheads]{llncs}

\usepackage{amsmath}
\usepackage{caption}
\usepackage{subcaption}
\usepackage{graphicx}
\usepackage{booktabs, makecell, tabularx}
\usepackage{rotating}
\usepackage[export]{adjustbox}
\usepackage{url}
\DeclareMathOperator*{\argmin}{\arg\!\min}

\newcommand{\norm}[1]{\left\lVert#1\right\rVert}

\begin{document}
\title{PCMC-T1: Free-breathing myocardial T1 mapping with Physically-Constrained Motion Correction\thanks{This research was supported in part by a grant from the United States-Israel Binational Science Foundation (BSF), Jerusalem, Israel.}}
\titlerunning{PCMC-T1: Free-breathing myocardial T1 mapping}
%

%
 \author{Eyal Hanania\inst{1}\orcidID{0000-0002-5268-8674} 
 \and Ilya Volovik\inst{2}\orcidID{0009-0000-0038-6597}
\and
Lilach Barkat\inst{3}\orcidID{0000-0003-3966-4798}
\and Israel Cohen\inst{1}\orcidID{0000-0002-2556-3972} 
\and
Moti Freiman\inst{3}\orcidID{0000-0003-1083-1548}}
\authorrunning{E. Hanania et al.}
\institute{Faculty of Electrical \& Computer Engineering, Technion - IIT, Haifa, Israel \and
Bnai Zion medical center, Haifa, Israel \and
Faculty of Biomedical Engineering, Technion - IIT, Haifa, Israel
\\
 \email{eyalhan@campus.technion.ac.il}}


\maketitle              
\begin{abstract}
$T_1$ mapping is a quantitative magnetic resonance imaging (qMRI) technique that has emerged as a valuable tool in the diagnosis of diffuse myocardial diseases.
However, prevailing approaches have relied heavily on breath-hold sequences to eliminate respiratory motion artifacts. This limitation hinders accessibility and effectiveness for patients who cannot tolerate breath-holding. Image registration can be used to enable free-breathing $T_1$ mapping. Yet, inherent intensity differences between the different time points make the registration task challenging. 
We introduce PCMC-T1, a physically-constrained deep-learning model for motion correction in free-breathing $T_1$ mapping. We incorporate the signal decay model into the network architecture to encourage physically-plausible deformations along the longitudinal relaxation axis. 
We compared PCMC-T1 to baseline deep-learning-based image registration approaches using a 5-fold experimental setup on a publicly available dataset of 210 patients. PCMC-T1 demonstrated superior model fitting quality ($R^2$: 0.955) and achieved the highest clinical impact (clinical score: 3.93) compared to baseline methods (0.941, 0.946 and 3.34, 3.62 respectively). Anatomical alignment results were comparable (Dice score: 0.9835 vs. 0.984, 0.988). 
Our code and trained models are available at \url{https://github.com/eyalhana/PCMC-T1}.

\end{abstract}

\section{Introduction}
Quantitative $T_1$ mapping is a magnetic resonance imaging (MRI) technique that allows for the precise measurement of intrinsic longitudinal relaxation time in myocardial tissue \cite{taylor2016t1}. 
``Native'' $T_1$ mapping, acquired without administration of a paramagnetic contrast agent, has been found to be sensitive to the presence of myocardial edema, iron overload, as well as myocardial infarcts and scarring \cite{schelbert2016state}.
It is increasingly recognized as an indispensable tool for the assessment of diffuse myocardial diseases such as diffuse myocardial inflammation, fibrosis, hypertrophy, and infiltration \cite{taylor2016t1}. 

\begin{figure}[t!]
\includegraphics[width=\textwidth]{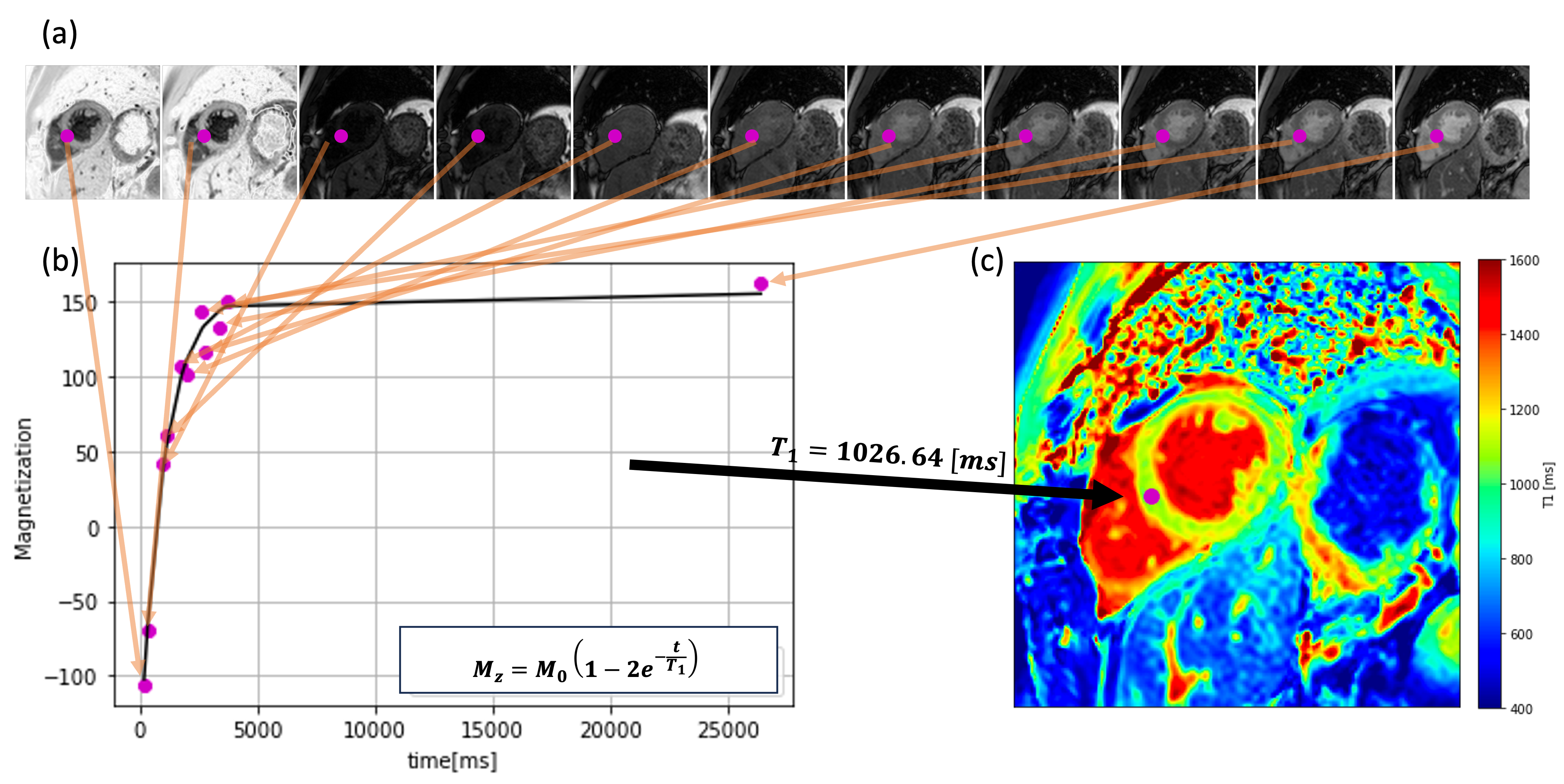}
\caption{Schematic description of $T_1$ mapping for a single voxel. (a) $T_1$-weighted myocardial images at 11 sequential time points. (b) Fitting an inversion recovery curve of the longitudinal magnetization $M_z$ over different time points $t$, and extracting the corresponding $T_1$ and $M_0$ parameters. (c) Displaying the computed $T_1$ map. }
\label{fig:relaxation}
\end{figure}

The derivation of accurate $T_1$ maps necessitates a sequential acquisition of registered images, where each pixel characterizes the same tissue at different timepoints (Fig.~\ref{fig:relaxation}). However, the inherent motion of the heart, respiration, and spontaneous patient movements can introduce substantial distortions in the $T_1$ maps, ultimately impeding their reliability and clinical utility, and potentially leading to an erroneous diagnosis. \cite{tilborghs2019robust}. 
Echo-triggering is a well-established approach to mitigate the effects of cardiac motion. Conversely, breath-hold sequences such as the Modified Look-Locker Inversion recovery (MOLLI) sequence and its variants \cite{roujol2014accuracy} are commonly employed to suppress motion artifacts associated with respiration. However, the requirement for subjects to hold their breath places practical constraints on the number of images that can be acquired \cite{roujol2014accuracy}, as well as on the viability of the technique for certain patient populations who cannot tolerate breath-holding. Further, inadequate echo-triggering due to cardiac arrhythmia may lead to unreliable $T_1$ maps, compromising the diagnosis.

Alignment of the images obtained at different time-points via image registration can serve as a mitigation for residual motion and enable cardiac $T_1$ mapping with free-breathing sequences such as the slice-interleaved $T_1$  (STONE) sequence \cite{weingartner2015free}. Yet, the intrinsic complexity of the image data, including contrast inversion, partial volume effects, and signal nulling for images acquired near the zero crossing of the $T_1$ relaxation curve, presents a daunting task in achieving registration for these images. 
Zhang et al. \cite{zhang2018cardiac} proposed to perform motion correction in $T_1$ mapping by maximizing the similarity of normalized gradient fields in order to address the intensity differences across different time points. El-Rewaidy et al. \cite{el2018nonrigid} employed a segmentation-based approach in which the residual motion was computed by matching manually annotated contours of the myocardium to the different images. Xue et al. \cite{xue2012motion} and Tilborghs et al. \cite{tilborghs2019robust} proposed an iterative approach in which the signal decay model parameters are estimated and synthetic images are generated. Then, image registration used the predicted images to register the acquired data. Van De Giessen et al. \cite{van2013model} used directly the error on the exponential curve fitting as the registration metric to spatially align images obtained from a Look-Locker sequence.

Deep-learning methods have been also proposed for motion correction by image registration as a pre-processing step in quantitative cardiac $T_1$ mapping \cite{gonzales2021moconet,li2022motion,arava2021deep}. A recent study by Yang et al. \cite{yang2022disq} introduced a sequential process to address the contrast differences between images. Initially, their approach aimed to separate intensity changes resulting from different inversion times from the fixed anatomical structure. However, this method heavily relied on the perfect disentanglement of the anatomical structure from the contrast. Moreover, the registration is performed exclusively between the disentangled anatomical images, overlooking the adherence of the signal along the inversion time axis to the signal decay model. Nevertheless, these methods do not account directly for the signal decay model, therefore they may produce physically-unlikely deformations. 


In this work, we introduce PCMC-T1, a physically-constrained deep-learning model for simultaneous motion correction and $T_1$ mapping from free-breathing acquisitions. Our network architecture combined an image registration module and an exponential $T_1$ signal decay model fitting module. The incorporation of the signal decay model into the network architecture encourages physically-plausible deformations along the longitudinal relaxation axis.  

Our PCMC-T1 model has the potential to expand the utilization of quantitative cardiac $T_1$  mapping to patient populations who cannot tolerate breath-holding by enabling automatic motion-robust accurate $T_1$  parameter estimation without additional manual annotation of the myocardium.

\section{Method}
We formulate the simultaneous motion correction and signal relaxation model estimation for qMRI $T_1$ mapping as follows:
\begin{equation}
\widehat{T_1,M_0,\Phi}= \argmin_{T_1,M_0,\Phi}\sum_{i=0}^{N-1}\norm{M_0\cdot(1-2\cdot e^\frac{-t_i}{T_1})-\phi_i\circ I_i}^2
\end{equation}
where $N$ is the number of acquired images,
$M_0, T_1$ are the exponential signal relaxation model parameters, $\phi_i$ is the i'th deformation field, $\Phi=\{\phi\}_{i=0}^{N-1}$, $I_i$ is the i'th original image, and $t_i$ is the i'th timestamp. However, direct optimization of this problem can be challenging and time-consuming \cite{van2013model}. 
\subsection{Model architecture}
\begin{figure}[t!]\includegraphics[width=0.98\textwidth]{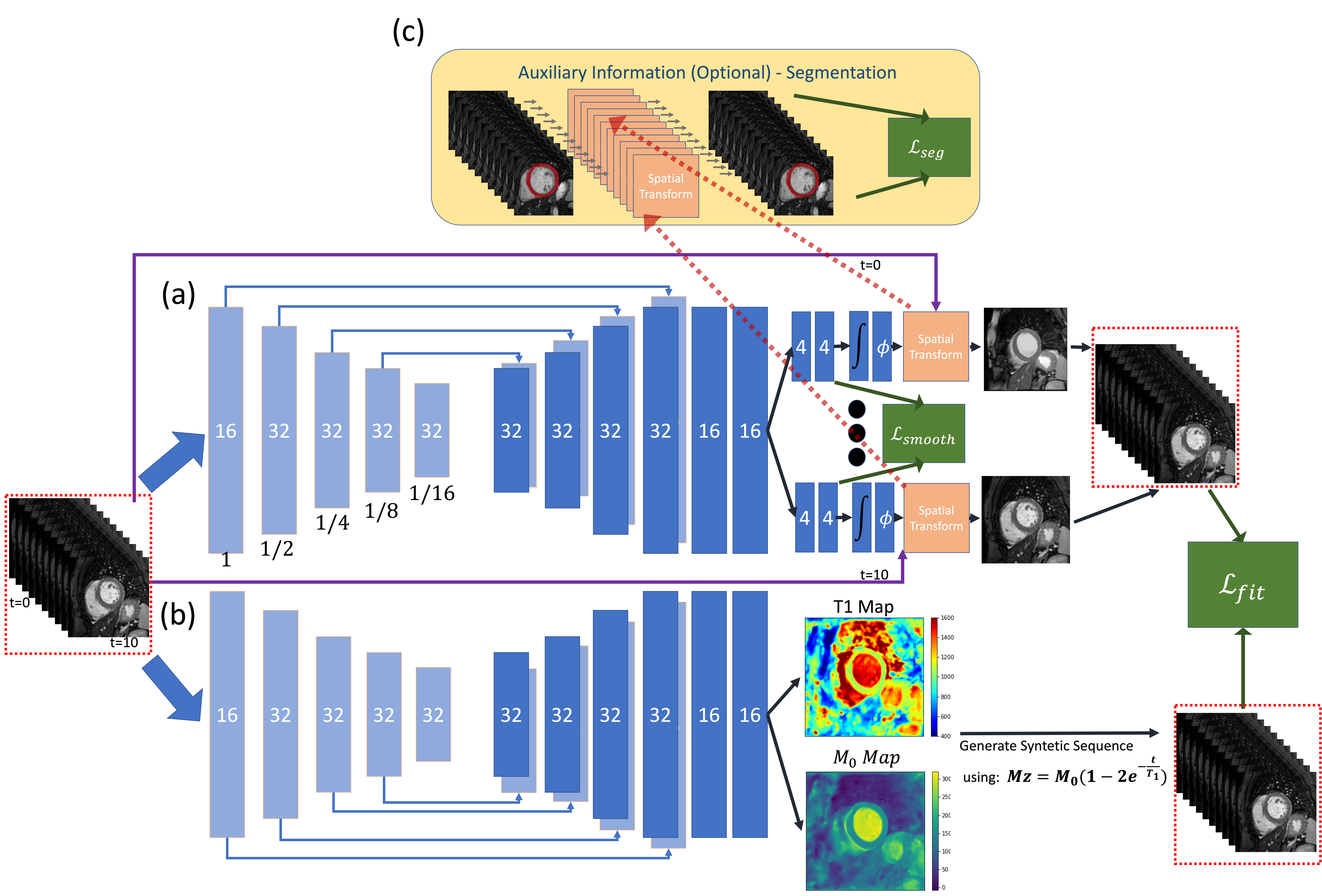}
\caption{Our PCMC-T1 model comprises two encoder-decoder components. (a) The first encoder-decoder extends the pair-wise VoxelMorph model to enable the registration of multiple images. (b) The second encoder-decoder generates parametric maps and motion-free synthetic images. The main goal of our network is to minimize the discrepancy between the registered images and the motion-free synthetic images, aiming for physically plausible deformations along the longitudinal relaxation axis. Additionally, we optionally promote anatomically consistent deformation fields by introducing a segmentation loss (c).}
\label{fig:network_architecture}
\end{figure}

To overcome this challenge, we propose PCMC-T1, a DNN architecture that simultaneously predicts the deformation fields and the exponential signal relaxation model parameters. 
Fig.~\ref{fig:network_architecture} summarizes the overall architecture of our model. It includes two U-Net-like encoder-decoder modules that are operating in parallel. Skip connections are connecting between the encoder and the decoder of each model. The first encoder-decoder module is a multi-image deformable image registration module based on the voxelmorph architecture \cite{balakrishnan2019voxelmorph}, while the second encoder-decoder module is the qMRI signal relaxation model parameters prediction module.
The input of the DNN is a set of acquired images $\{I_i|i=0 \ldots N-1\}$, stacked along the channel dimension. 
The first encoder-decoder is an extension of the pair-wise VoxelMorph model \cite{balakrishnan2019voxelmorph} for registration of multiple images. The encoder is a U-Net-like encoder consisting of convolutional and downsampling layers with an increasing number of filters. The decoder output splits into multiple separated heads of convolutional layers and integration layers that produce a specific deformation field $\{\phi_i|i=0 \ldots N-1\}$ for each timestamp $i$. Skip connections are used to propagate the learned features into the deformation field prediction layers. A spatial warping layer is used to align the acquired images $I_i$ to the synthetics images generated from the signal relaxation model parameters predictions ($S_i$):  $R_i=I_i\circ\phi_i$. The specific details of the architecture are as in Balakrishnan et al \cite{balakrishnan2019voxelmorph} and the details of the integration layer are as in Dalca et al \cite{dalca2019unsupervised}.   

The second encoder-decoder has a similar architecture. It has two output layers representing the exponential signal relaxation model parameters: $T_1$ and $M_0$. The predicted parameters maps are then used, along with the input's timestamps $\{t_i|i=0 \ldots N-1\}$,  as input to a signal generation layer. This layer generates a set of motion-free images  $\{S_i|i=0 \ldots N-1\}$ computed directly from the estimated parametric maps ($M_0$, $T_1$) at the different inversion times using the signal relaxation model \cite{weingartner2015free}:
\begin{align}
\label{eq:signal_model}
S_i = M_0\cdot(1-2\cdot e^\frac{-t_i}{T1})
\end{align}

\subsection{Loss Functions}
We encourage predictions of physically-plausible deformation fields by coupling three terms in our loss function as follows: 
\begin{equation}
\mathcal{L}_{total} = \lambda_1\cdot \mathcal{L}_{fit} + \lambda_2\cdot \mathcal{L}_{smooth} + \lambda_3\cdot \mathcal{L}_{seg}
\end{equation}
The first term ($\mathcal{L}_{fit}$) penalizes for differences between the model-predicted images generated by the model-prediction decoder and the acquired images warped according to the deformation fields predicted by the registration decoder. Specifically, we use the mean-squared-error (MSE) between the registered images $\{R_i|i=0 \ldots N-1\}$ and the synthetic images $\{S_i|i=0 \ldots N-1\}$: 
\begin{align}
\mathcal{L}_{fit}(T_1,M_0,t_{i=0}^{N-1},\Phi) = \sum_{i=0}^{N-1}(S_i - R_i)^2 
\end{align}
where $S_i$ are the images generated with the signal model equation (Eq.~\ref{eq:signal_model}), and the registered images are the output of the registration module. This term encourages deformation fields that are physically plausible by means of a signal relaxation that is consistent with the physical model of $T_1$ signal relaxation. 

The second term ($\mathcal{L}_{smooth}$) encourages the model to predict  realistic, smooth deformation fields $\Phi$ by penalizing for a large $l_2$ norm of the gradients of the velocity fields \cite{balakrishnan2019voxelmorph}: 
\begin{align}
\mathcal{L}_{smooth}(\Phi) =\sum_{i=0}^{N-1}\frac{1}{\Omega} \sum_{p \in \Omega}{\norm{\nabla\phi_i(p)}}^2
\end{align}
where $\Omega$ is the domain of the velocity field and $p$ are the voxel locations within the velocity field. 
In addition, we encourage anatomically-consistent deformation fields by introducing a segmentation-based loss term ($\mathcal{L}_{seg}$)as a third term in the overall loss function \cite{balakrishnan2019voxelmorph}. This term can be used in cases where the left ventricle (LV)'s epicardial and endocardial contours are available during training. Specifically, the segmentation loss function is defined as follows:
\begin{equation}
\mathcal{L}_{seg}(r,Seg_{i=0}^{N-1},\phi_{i=0}^{N-1}) = \sum_{i=0,i\neq r}^{N-1} DiceLoss(Seg_r,Seg_i \circ \phi_i)
\end{equation}
where $Seg_i, (i \in {0,\ldots,N-1})$ is the $i'th$ binary segmentation mask of the myocardium, $Seg_r$ is the binary segmentation mask of the fixed image, and $r$ is the index of the fixed image. This term can be omitted in cases where the segmentations of the myocardium are not available.
\subsection{Implementation details}
We implemented our models in PyTorch. We experimentally fixed the first time-point image, and predict deformation fields only for the rest of the time points. We optimized our hyperparameters using a grid search. The final setting for the loss function parameters were: $\lambda_1$ = 1, $\lambda_2$ = 500, $\lambda_3$ = 70000. We used a batch size of 8, ADAM optimizer with a learning rate of $2\cdot10^{-3}$. We trained the model for 300k iterations.  We used the publicly available TensorFlow implementations of the diffeomorphic VoxelMorph \cite{dalca2019unsupervised} and SynthMorph \cite{hoffmann2021synthmorph} as baseline methods for comparison. We performed hyper-parameter optimization for baseline methods using a grid search. All experiments were run on an NVIDIA Tesla V100 GPU with 32G RAM.

\section{Experiments and results}
\subsection{Data}
We used the publicly available myocardial $T_1$ mapping dataset \cite{el2018nonrigid,cardiacdataset}. The dataset includes 210 subjects, 134 males and 76 females aged $57\pm14$ years, with known or suspected cardiovascular diseases. The images were acquired with a 1.5T MRI scanner (Philips Achieva) and a  32-channel cardiac coil using the ECG-triggered free-breathing imaging slice-interleaved $T_1$ mapping sequence (STONE) \cite{weingartner2015free}. Acquisition parameters were: field of view (FOV) $=360\times351[mm^2]$, and voxel size of $2.1\times2.1\times8[mm^3]$. For each patient, 5 slices were acquired from base to apex in the short axis view at 11 time points. Additionally, manual expert segmentations of the myocardium were provided as part of the dataset \cite{el2018nonrigid}. 
We cropped the images to a size of $160\times160$ pixels for each time point. We normalized the images using a min-max normalization.

\subsection{Evaluation Methodology}
\begin{table}[t]
\centering
\caption{Quantitative comparison between motion correction methods for myocardial $T_1$ mapping. All results are presented in mean$\pm$std.}
\footnotesize
\begin{tabular}{|c|c|c|c|c|c|c|}
\hline
\textbf{}                                             
& $R^2$    & $DSC$  & HD [mm] & clinical \\
&          &        &          &score  \\ \hline
Original                                         & $0.911\pm0.12$       & $0.664\pm0.23$           & $14.93\pm11.76$         & $2.79\pm0.99$     \\
SynthMorph                                         &$0.946\pm0.09 $        &$0.88\pm0.149  $             &$8.59\pm9.98$      & $3.62\pm0.88$                    \\
Voxelmorph-seg                                    & $0.941\pm0.096$     & $0.84\pm0.188$            & $9.39\pm11.93$         & $3.34\pm0.79$         \\
Reg-MI                                          & $0.95\pm0.08$         & $0.73\pm 0.168 $           & $16.29\pm11.43$                &  $3.68\pm0.83$           \\
PCMC-T1 w.o $\mathcal{L}_{seg}$                &$0.971\pm0.046$            &$0.662\pm0.172$              & $21.5\pm13.3$ &          $4\pm0.83$                         \\
PCMC-T1                                         & $0.955\pm0.078$      & $0.835\pm0.137$           & $9.34\pm7.85$            & $3.93\pm0.78$   \\
\hline
\end{tabular}
\label{table:results}
\end{table}
{\bf Quantitative evaluation:} We used a 5-fold experimental setup. For each fold, we divided the 210 subjects into 80\% as a training set and 20\% as a test set. We conducted an ablation study to determine the added value of the different components in our model. Specifically, we compared our method using a few variations, including a multi-image registration model with a mutual-information-based loss function (REG-MI)\cite{hanania2023groupT1}, and our method (PCMC-T1) without the segmentation loss term. We used two state-of-the-art deep-learning algorithms for medical image registration including the pairwise probabilistic diffeomorphic VoxelMorph with a mutual-information-based loss \cite{dalca2019unsupervised}, and pairwise SynthMorph \cite{hoffmann2021synthmorph}, as well as with $T_1$ maps produced from the acquired images directly without any motion correction step.  We quantitatively evaluated the $T_1$ maps produced by our PCMC-T1 model in comparison to $T_1$ maps produced after applying deep-learning-based image registration as a pre-processing step. We used the $R^2$ of the model fit to the observed data in the myocardium, the Dice score, and Hausdorff distance values of the myocardium segmentations as the evaluation metrics. 

\noindent{\bf Clinical impact:} We further assessed the clinical impact of our method by conducting a semi-quantitative ranking of the $T_1$ maps for the presence of motion artifacts by an expert cardiac MRI radiologist (3 years of experience) who was blinded to the methods used to generate the maps. We randomly selected 29 cases (5 slices per case) from the test set with their associated $T_1$ maps. The radiologist was asked to rank each slice with 1 in case of a good quality map without visible motion artifacts and with 0 otherwise. We computed overall patient scores by summing the slice grades. The maximum grade per subject was 5 for cases in which no motion artifacts were present in all slices and 0 for cases in which motion artifacts were present in all slices. We assessed the statistical significance with the repeated measures ANOVA test; p$<$0.05 was considered significant.
\subsection{Results}
{\bf Quantitative evaluation:} Table~\ref{table:results} summarizes our results for the test sets across all folds, encompassing a total of 210 patients. Our PCMC-T1 approach achieved the best result in terms of $R^2$ with the smallest variance. Although PCMC-T1 without the segmentation loss (${L}_{seg}$) achieved a higher $R^2$ result compared to PCMC-T1 with the segmentation loss, it degraded the Dice value, representing over-fitted predictions. On the other hand, the slightly higher Dice score and Hausdorff distance values obtained by baseline methods compared to PCMC-T1 suggest bias of these methods toward the registration of the segmentation maps rather than producing deformation fields that are consistent with the signal relaxation model. The balanced result of PCMC-T1 indicates an improvement in the physical plausibility of the deformations produced by PCMC-T1 by means of signal relaxation and anatomical consistency. 

\begin{figure}[t!]
\begin{tabularx}{\linewidth}{@{\hskip -40pt}c@{\hskip -56pt}c@{\hskip -56pt}c@{\hskip -56pt}c}
{\bf Original} & {\bf VoxelMorph} & {\bf SynthMorph} & {\bf PCMC-T1} \\
\includegraphics[width=0.42\textwidth]{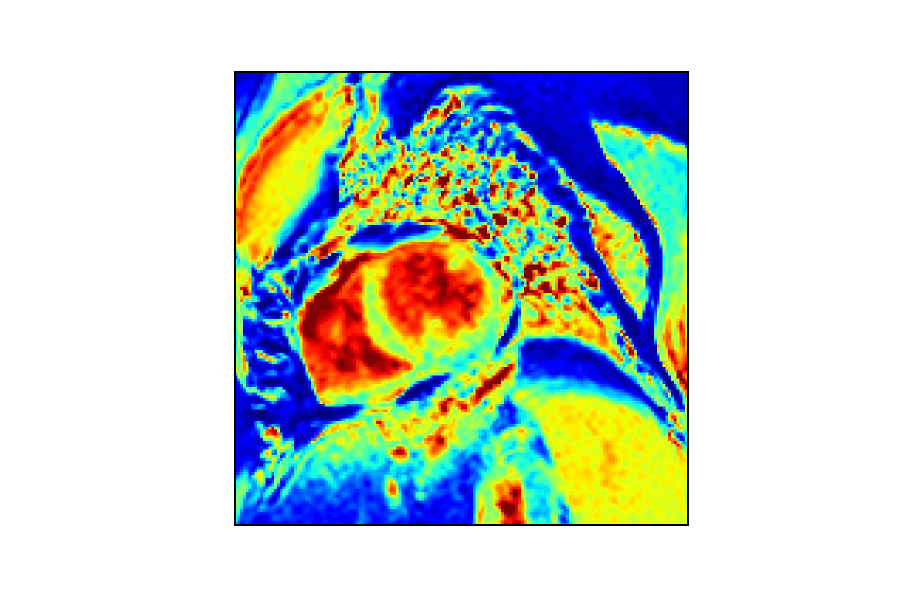} &   \includegraphics[width=0.42\textwidth]{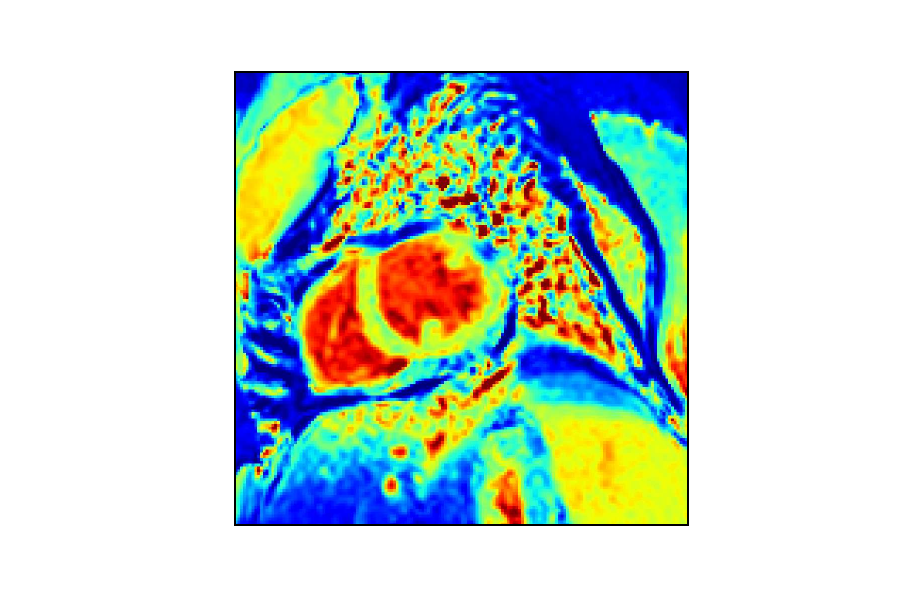} &   \includegraphics[width=0.42\textwidth]{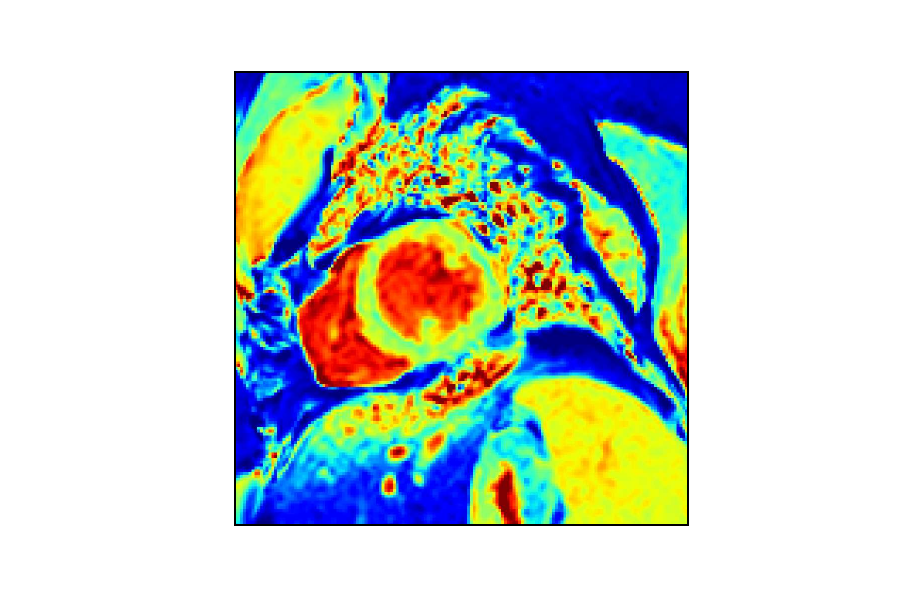} &
\includegraphics[width=0.42\textwidth]{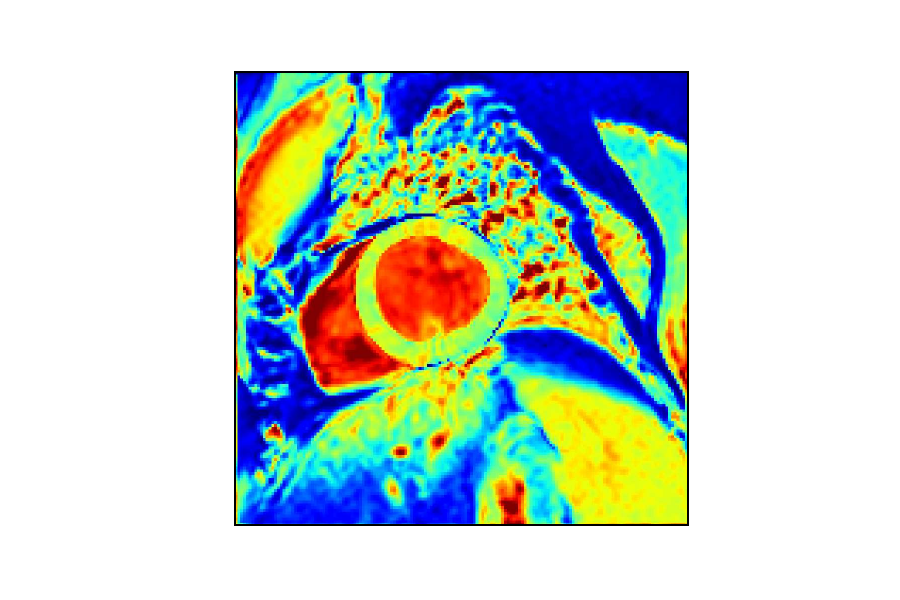} 
\vspace{-12pt}
\\   
\includegraphics[width=0.42\textwidth]{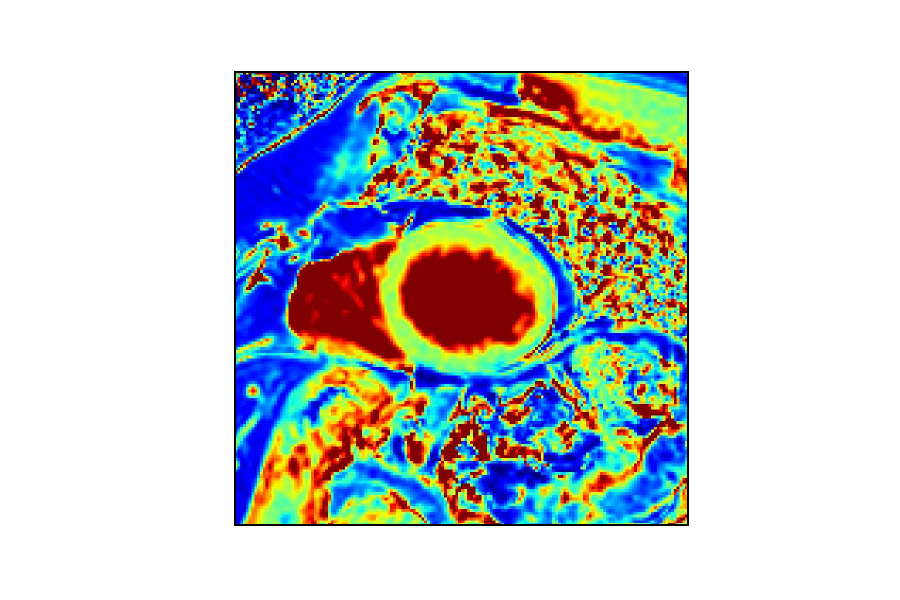} &   \includegraphics[width=0.42\textwidth]{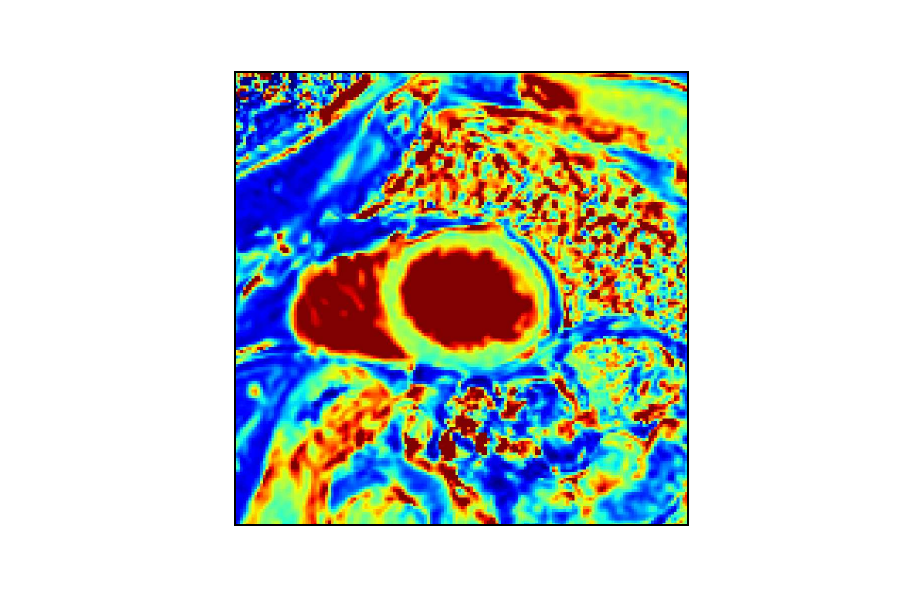} &   \includegraphics[width=0.42\textwidth]{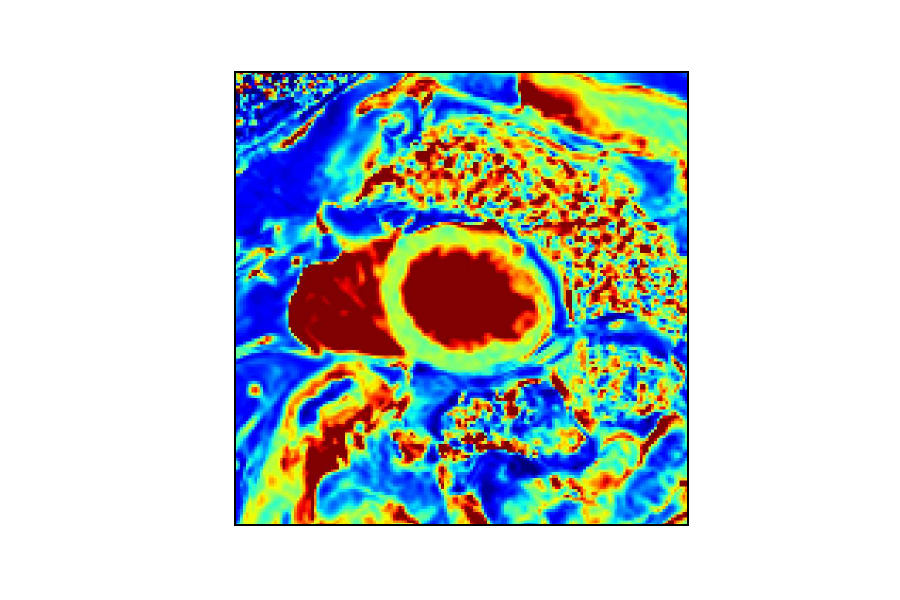} &
\includegraphics[width=0.42\textwidth]{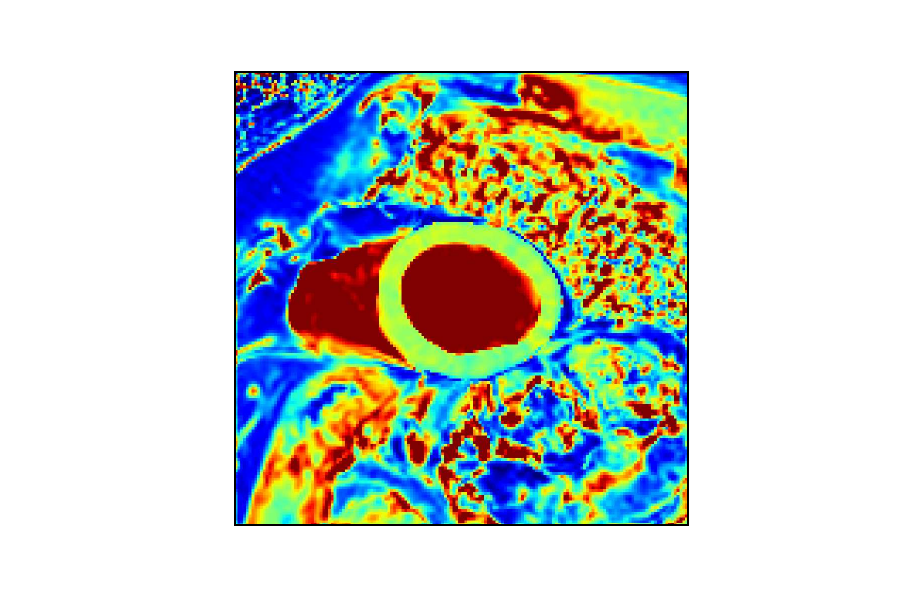} \vspace{-12pt}
\\  
\includegraphics[width=0.42\textwidth]{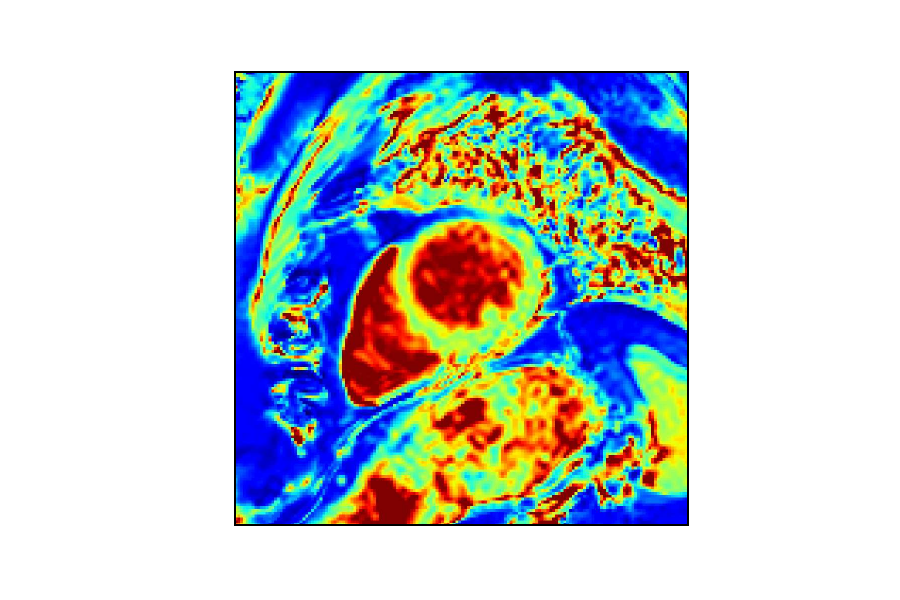} &   \includegraphics[width=0.42\textwidth]{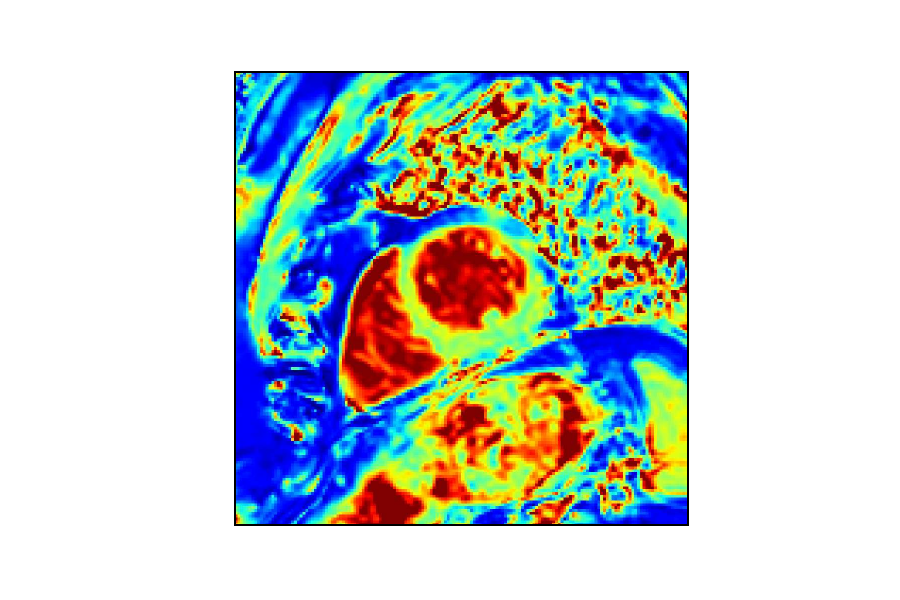} &   \includegraphics[width=0.42\textwidth]{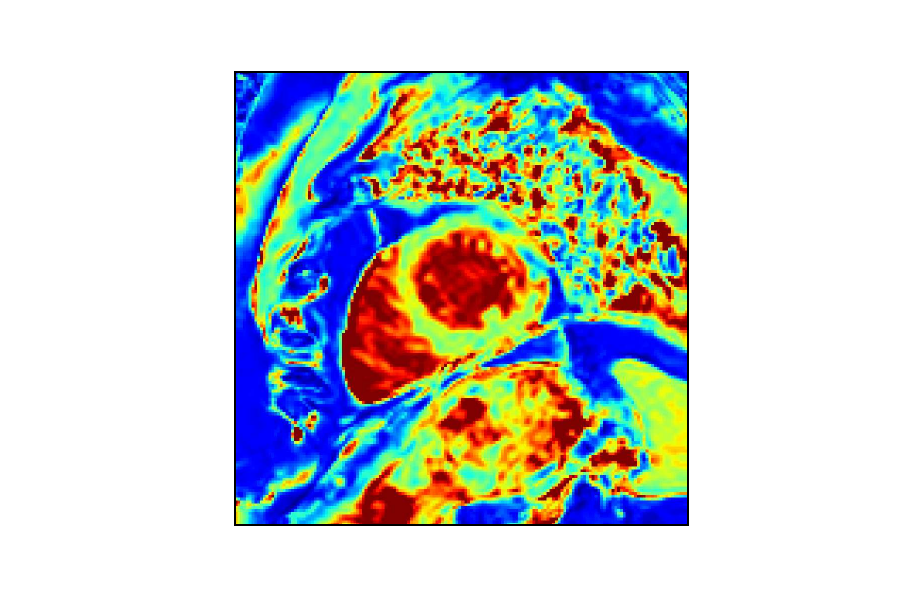} &
\includegraphics[width=0.42\textwidth]{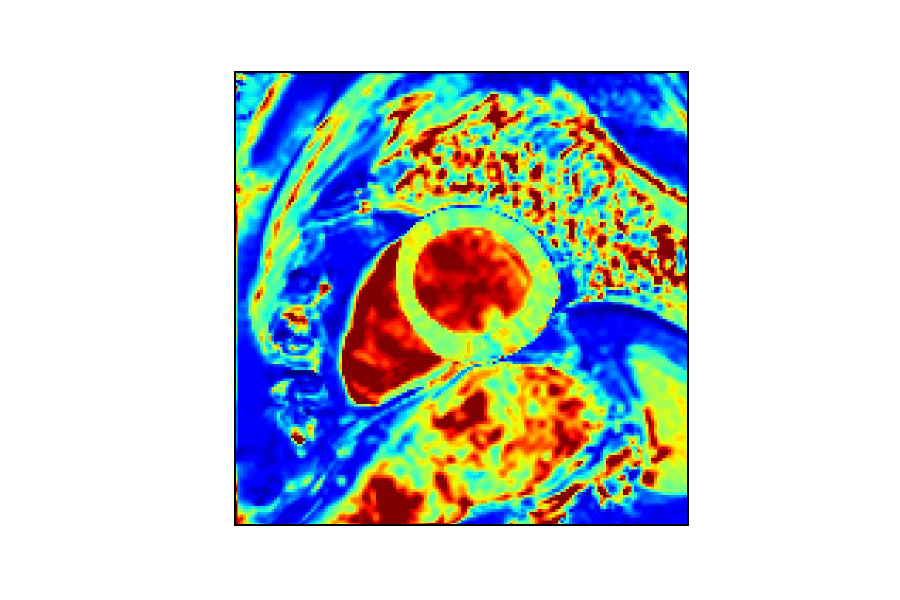} 
\end{tabularx}
\vspace{-12pt}
\caption{Representative $T_1$ maps computed with the different approaches. Our approach (PCMC-T1) demonstrates a clearer delineation between the blood and the muscle with a reduced partial volume effect, resulting in a more homogeneous mapping of the myocardium.}
\label{fig:clinical}
\end{figure}

\noindent{\bf Clinical impact:}
Fig.~\ref{fig:clinical} presents several representative cases. Although the Dice score of the baseline methods is higher compared to this of PCMC-T1, the quality of the maps produced by PCMC-T1 is better. The rightmost column of Table~\ref{table:results} summarizes the results of the clinical impact assessment of our PCMC-T1 approach. Our PCMC-T1 received the highest quality score compared to the baseline methods. The difference in the radiologist grading was statistically significant (p$ \ll 10e^{-5}$). The improvement in the radiological evaluation suggests that PCMC-T1 provides a balanced result that is not overly biased toward the segmentations or toward the signal relaxation model.
\section{Conclusions}
We presented PCMC-T1, a physically-constrained deep-learning model for motion correction in free-breathing $T_1$ mapping. Our main contribution is the incorporation of the signal decay model into the network architecture to encourage physically-plausible deformations along the longitudinal relaxation axis. We demonstrated a quantitative improvement by means of fit quality with comparable Dice score and Hausdorff distance. We further assessed the clinical impact of our method by conducting a qualitative evaluation of the $T_1$ maps produced by our method in comparison to baseline methods by an expert cardiac radiologist.  
Our PCMC-T1 model holds the potential to broaden the application of quantitative cardiac $T_1$ mapping to patient populations who are unable to undergo breath-holding MRI acquisitions by enabling motion-robust accurate $T_1$ parameter estimation. Further, the proposed physically-constrained motion robust parameter estimation approach can be directly extended to quantitative T2 mapping as well as to additional qMRI applications.


\bibliographystyle{splncs04}
\bibliography{bibliography}

\end{document}